\documentstyle[12pt]{article}
\def\bib{\bibitem}
\def\be{\begin{equation}}
\def\ee{\end{equation}}
\def\barr{\begin{array}}
\def\earr{\end{array}}

\def\etal{ {\it et al.} }
\def\lsim{\:\raisebox{-0.5ex}{$\stackrel{\textstyle<}{\sim}$}\:}

\def\rp{$R_p \hspace{-1em}/\;\:$}
\begin{document}
\setcounter{page}{0}
\renewcommand{\thefootnote}{\fnsymbol{footnote}}
\thispagestyle{empty}
\vspace*{-1in}
\begin{flushright}
CERN-TH/95--54 \\[2ex]
{\large \bf hep-ph/9503263} \\
\end{flushright}
\vskip 45pt
\begin{center}
{\Large \bf \boldmath $D$- and $\tau$-Decays: Placing
New Bounds on
$R$-Parity--Violating Supersymmetric Couplings}

\vspace{11mm}
\bf
Gautam Bhattacharyya\footnote{ gautam@cernvm.cern.ch} and
Debajyoti Choudhury\footnote{debchou@surya11.cern.ch}\\

\vspace{13pt}
{\bf Theory Division, CERN, \\ CH--1211 Gen\`eve 23, Switzerland.}

\vspace{50pt}
{\bf ABSTRACT}
\end{center}

\begin{quotation}

$D$- and $\tau$-decays are used to place bounds on some
$R$-parity--violating $\lambda'$-type Yukawa interactions. Some
of these bounds are competitive with the existing ones, some are
improved while some are new.

\end{quotation}

\vspace{170pt}
\noindent
\begin{flushleft}
CERN-TH/95--54\\
March 1995\\
\end{flushleft}

\centerline{(Submitted to {\em Mod. Phys. Lett.} {\bf A})}

\vfill
\newpage
\setcounter{footnote}{0}
\renewcommand{\thefootnote}{\arabic{footnote}}

\setcounter{page}{1}
\pagestyle{plain}
\advance \parskip by 10pt

Of the ideas that take us beyond the standard model (SM),
supersymmetry is perhaps the most extensively discussed.
As the name suggests, the Minimal Supersymmetric Standard Model
(MSSM)
is obtained by the naive supersymmetrization of both the SM
particle content and the couplings \cite{mssm}.
Furthermore, an additional
Higgs supermultiplet has to be included both for anomaly
cancellation as well as for
fermion mass generation. A new feature
arises at this juncture. Since the $SU(2)$-doublet lepton
superfields  have the
same gauge quantum numbers as one of the higgs supermultiplets,
the latter can be replaced by the former in any or all of the Yukawa
interaction terms. One may also write trilinear terms involving the
$SU(2)$--singlet quark supermultiplets.
The additional pieces in the superpotential may thus
be parametrized as \cite{rpar}
\be
    {\cal W}_{\not R} =  \lambda_{ijk} L_i L_j E^c_k
                        +  \lambda'_{ijk} L_i Q_j D^c_k
                        +  \lambda''_{ijk} U^c_i D^c_j D^c_k \ ,
      \label{R-parity}
\ee
where $L_i$ and $Q_i $ are the $SU(2)$-doublet lepton and quark
superfields and $E^c_i, U^c_i, D^c_i$
are the singlet  superfields. Clearly $\lambda_{ijk}$ is antisymmetric
under the interchange of the first two indices, while $\lambda''_{ijk}$
is antisymmetric under the interchange of the last two.

It is obvious that the presence of such terms can alter
phenomenology to a great degree. For example, while the first two
terms in eq.(\ref{R-parity}) violate lepton number, the last one
violates baryon number. The simultaneous presence of both sets
can, therefore,
lead to a catastrophically high rate for proton decay.
This and other
such issues have provoked the introduction of a discrete symmetry
known as ``matter parity'' or equivalently,
``$R$--parity''. Representable as
$R = (-1)^{3B + L + 2 S}$, where $B,L,S$ are the baryon number, lepton
number and the intrinsic spin of the field respectively,
$R$ has a value of $+1$ for
all SM particles and $-1$ for all their superpartners.
This symmetry rules out each of the terms in eq.(\ref{R-parity}),
with the additional
consequence that the lightest superpartner (LSP) must be stable.

While an exact $R$-parity is a sufficient condition for the
suppression
of certain unobserved processes, the theoretical motivation for such
a symmetry is not clear. This makes the question of establishing
phenomenological bounds on $R$-violating couplings an interesting one.
This issue is of paramount importance in the context of the search
for supersymmetric particles in the forthcoming colliders. Even a
tiny $R$-parity--violating (\rp) coupling can totally change
the expected signatures.

The constraints imposed by the non--observance of proton decay
can be  circumvented by
assuming that all of $\lambda''_{ijk}$ are zero. Such
a scenario might be motivated within certain theoretical
frameworks \cite{hall-suzuki}, and we implicitly assume the same
 in the
rest of this note. This assumption also renders simpler the
problem of preservation of GUT--scale baryogenesis \cite{baryo}.
Although the presence of the other \rp\  terms can
also affect the baryon asymmetry of the universe,
Dreiner and Ross \cite{dreiner} have argued
that such bounds are highly model-dependent and  can
hence be evaded. For example,
in cases where at least one $L$-violating
coupling involving a particular lepton
family is small enough ($ \lsim 10^{-7}$)
so as to (almost) conserve the corresponding lepton flavour
over cosmological time scales, such bounds are no longer effective.

In what follows, we focus our attention only on the $\lambda'$-type
$L$-violating Yukawa interactions.
Bounds on individual $\lambda'$-type couplings have been
derived from limits on the Majorana mass of $\nu_e$, by
demanding charge--current universality in various decays
and also from analyses
of atomic parity violation, forward--backward  asymmetry in
$e^+ e^-$ collisions, deep inelastic scattering {\em etc.}
A comprehensive
study can be found in refs.\cite{barger}.
However, while many of the $\lambda'$-type
couplings are constrained to be  $\lsim 0.1$ for a common
sfermion mass ($\tilde{m}$) of 100 GeV,
the existing low--energy bounds on some of them are still relatively
weak. Additionally,
LEP data on the $Zl\bar{l}$ couplings have been used to derive some
bounds on $\lambda'_{i3k}$ ($\sim 0.5$ for $\tilde{m} = 100$ GeV) \cite{gg}.

As far as  $\lambda'_{i2k}$ are concerned,  constraints
exist for only some of them  and these  are not very stringent
either. Derived mainly from three classes
of experiments : ($i$) $\nu_\mu$-induced deep inelastic scattering,
($ii$) forward--backward  asymmetry in
$e^+ e^-$ collisions and ($iii$) atomic parity violation and
$eD$ asymmetry, the bounds range between $0.22$--$0.45$
(for $\tilde{m} = 100$ GeV) at the $1 \sigma$ level.
It has been suggested in ref.\cite{grt} though
that the presence of $\lambda'_{1jk}$
would induce, at the one--loop level, a Majorana mass for $\nu_e$.
The upper bound on the latter could then be used to place stringent limits
on $\lambda'_{1jk}$ (and similarly on $\lambda_{1jk}$ too). Such
arguments, if correct,
 would also hold, to a lesser extent, for $\lambda_{2jk}$ and
$\lambda'_{2jk}$ as well.
However, a look at the Lagrangian in eqns.(\ref{R-parity},\ref{lambda-pr}),
shows that this  constraint is applicable {\em only} for
$j = k$, and {\em not} for the general case as suggested in
ref.\cite{grt}. The only relevant bound from $\nu_e$ mass
on $\lambda'_{i2k}$
is then that on  $\lambda'_{122}$ which is now
constrained to be  $\lsim 0.04$ (at $1 \sigma$)
for $\tilde{m} = 100$ GeV.
Bounds on $\lambda'_{31k}$ and
$\lambda'_{32k}$ are, as yet, non-existent.

In this short note
we attempt to improve the above situation.
On the one hand,  experimental data on the observed decays of $D$-mesons
are utilized to place bounds on $\lambda'_{i2k}$,
 on the other, $\tau$-decays
are used to constrain $\lambda'_{31k}$.
While some of these are new too, the others are at least
comparable to those existing in the literature,
and, in most cases, supplant them.

To keep the discussion simple, we
shall confine ourselves to semi-leptonic decays, and there too
to final states containing only a single meson.
{}From eq.(\ref{R-parity}), the relevant part of the
Lagrangian can be written (in terms of the component fields)
 as
\be
\barr{rcl}
   {\cal L}_{\lambda'} & = & \lambda'_{ijk}
            \left[ \overline{d_{kR} } \nu_{iL} \tilde{d}_{jL} +
                   \overline{d_{kR} } d_{jL} \tilde{\nu}_{iL} +
                   \overline{(\nu_{iL})^c } d_{jL} \tilde{d}^\ast_{kR}
            \right. \\[1.5ex]
& & \left. \hspace*{1.5em}
                   - \overline{d_{kR} } e_{iL} \tilde{u}_{jL} -
                   \overline{d_{kR} } u_{jL} \tilde{e}_{iL} -
                   \overline{(e_{iL})^c } u_{jL} \tilde{d}^\ast_{kR}
            \right] + {\rm h.c.}
\earr
      \label{lambda-pr}
\ee
As is evident,  the above \rp\ couplings manifest
themselves only when the two non--spectator quarks
 form an $SU(2)_L$ doublet.
 As an explicit example, we discuss the particular  decay
$D^+\:(c\bar{d}) \rightarrow \bar{K^0}(s\bar{d})\: \mu^+ \nu_\mu$.
The presence of the above interaction
introduces an additional
quark level diagram (involving the exchange of a squark,
say $\tilde{b}_R$ as a particular case) having the current structure
\be
\displaystyle
   \frac{\lambda^{\prime 2}_{223} }{m^2_{\tilde{b}_R} }\:
     \overline{(\mu_L)^c} c_L \;\: \overline{s_L} (\nu_{\mu L})^c\; .
      \label{rp-current}
\ee
A  Fierz reordering in eq.(\ref{rp-current})
takes it back to the SM current structure and, adding the two
contributions, the
effective current looks like
\be
\displaystyle
   \frac{1}{8} \left[ \frac{g^2}{m_W^2} V_{cs} +
       \frac{\lambda^{\prime 2}_{223} }{m^2_{\tilde{b}_R}} \right] \:
     \overline{\nu_\mu} \gamma_\rho (1 - \gamma_5) \mu \;\:
     \overline{s} \gamma^\rho (1 - \gamma_5) c\ .
      \label{effective}
\ee
It is easy to see that at the quark level, all the
decays (whether of mesons or of the $\tau$)
meeting the above-mentioned criteria
can be described by an effective four--fermion interaction
similar to that in eq.(\ref{effective}).
There now remains to
calculate the branching fractions and compare these to the experimental
results.
However, there exists a small complication
 in the case of $D$-decays.
The hadronic matrix elements
involved in these decays can be parametrized in terms of a few
form factors (two for $D \to Kl\nu$ and four for $D \to K^*l\nu$).
Although
these may be calculated to some degree of accuracy in various
models \cite{bsw}, the theoretical
errors involved are still somewhat
large. The straightforward approach of using each decay channel
separately should thus be avoided till these
form factors are  known better, say from calculations on the
lattice \cite{lattice}.
A better method, in this case, is to compare
the partial widths into electron and muon channel
respectively (keeping the mesons  the same). Since the
lepton masses are negligible compared
to the scale of the problem, we may safely ignore the
$q^2$--evolution of the form factors. These
no longer  appear in the ratio of the experimental
widths which thus gives a direct bound on the relative
deviation due to the \rp\ ~interaction.
Although the process of combining observables involves compounding
different experimental errors, this is more than offset
by the deliverance from the  theoretical uncertainties.
One should also bear in mind the
possibility that the sizes of the \rp~couplings in the
electronic and the muonic sector are similar and hence the
effects might cancel each other.
However, the data on FCNC processes already constrain the
products of different couplings to be very small. We shall then
work under the (reasonable) assumption that at most one
of these couplings is non--zero.
For $\tau$-decays, no such considerations are necessary.
As the pion decay constant $f_\pi$ is relatively well-measured,
we may compare the theoretical expression for the decay
$\tau \to \pi \nu_\tau$ with the experimentally determined partial
width to obtain our bounds on $\lambda'_{31k}$.

For the numerical computation we use the following
experimental inputs \cite{pdg}
\be
\barr{rrclrcl}
a)& \displaystyle
  \frac{ Br(D^+ \rightarrow \bar{K}^0 \mu^+ \nu_\mu ) }
        { Br(D^+ \rightarrow \bar{K}^0 e^+ \nu_e ) }
    & = & 1.06^{+ 0.48}_{- 0.34} \; ; \\[3ex]
b)& \displaystyle
 \frac{ Br(D^+ \rightarrow \bar{K}^{0 \ast} \mu^+ \nu_\mu ) }
        { Br(D^+ \rightarrow \bar{K}^{0 \ast} e^+ \nu_e ) }
    & = & 0.94 \pm 0.16 \; ;
     \\[3ex]
c)& \displaystyle
 \frac{ Br(D^0 \rightarrow K^- \mu^+ \nu_\mu ) }
        { Br(D^+ \rightarrow K^- e^+ \nu_e ) }
    & = & 0.84 \pm 0.12 \; ;
\earr
    \label{branching}
\ee
and
\be
d)~~~~ Br(\tau^- \rightarrow \pi^- \nu_\tau)  = 0.117 \pm 0.004,
\qquad
f_\pi = (130.7 \pm 0.1 \pm 0.36) \; {\rm MeV}.
   \label{fpi}
\ee
The constraints that we derive are summarized in Table 1.
The bounds on $\lambda'_{121}$ and $\lambda'_{123}$ are
already quite competitive\footnote{We also obtain similar numbers for
$\lambda'_{122}$, but the latter is
already constrained tightly from the upper bound on
$\nu_e$ Majorana mass.} with the ones existing in the
literature.
These are going to improve even further
once the $D$-branching fractions
are known to a better accuracy.
On the other hand,
for  $\lambda'_{221}$, our numbers
are already better than the existing constraints.
On the five couplings, $\lambda'_{222},
\lambda'_{223}$ and all $\lambda'_{31k}$, our method places
phenomenological bounds for the first time.
\begin{table}[htb]
\begin{center}
\bigskip
\begin{tabular}{|c|c|c|c|}
\hline
$\{ijk\}$ & Existing bounds & \multicolumn{2}{| c |}{Our bounds} \\[1ex]
\cline{3-4}
&& ($1 \sigma$) & ($2 \sigma$) \\
\hline
&&&\\[.2ex]
121,~123 & 0.26 \ ($1 \sigma$) & $0.30^{a)}$,~$0.28^{b)}$,~$0.31^{c)}$
                              & $0.45^{a)}$,~$0.36^{b)}$,~$0.38^{c)}$ \\
\hline
&&&\\[.5ex]
221      & 0.22 \ ($2 \sigma$) & $0.42^{a)}$,~$0.18^{b)}$
                                & $0.58^{a)}$,~$0.30^{b)}$,~$0.17^{c)}$
                                \\
222,~223 &      & $0.42^{a)}$,~$0.18^{b)}$
                                & $0.58^{a)}$,~$0.30^{b)}$,~$0.17^{c)}$
                                \\
\hline
&&&\\[.5ex]
$31k$      &      &  \multicolumn{1}{|c|}{ $0.14^{d)}$ } & $0.18^{d)}$ \\
\hline
\end{tabular}
\caption[] {The  upper bounds on the $\lambda'$--type \rp~ couplings
(for $\tilde{m} = 100$ GeV)
obtained from our analyses of $D$- and $\tau$-decays.
The specific processes are labelled by the superscripts, $a$) to
$d$), exactly as they correspond to the experimental inputs shown
in eqs.(\protect\ref{branching}--\protect\ref{fpi}).
At the $1 \sigma$ level, there are no bounds on $\lambda'_{22k}$
from process ($c$).
The existing bounds are quoted from ref.\protect\cite{barger}.
}
\label{table}
\end{center}
\end{table}

To summarise, we use the data on $D$- and $\tau$-decays to
constrain some of the $\lambda'$-type
\rp\ couplings within the
MSSM. $D$-decays constrain $\lambda'_{12k}$
and $\lambda'_{22k}$, while $\tau$-decay constrains $\lambda'_{31k}$.
Bounds obtained on $\lambda'_{121}$ and $\lambda'_{123}$ are
{\em at par} with the existing ones,
while those obtained on  $\lambda'_{221}$
are already {\em better}.
Further improvements can be expected from two sources :
($i$) an increase in the experimental accuracies,
and/or ($ii$) a better determination of
form factors involved in $D$-decays  enabling us to use each decay
channel as an independent input rather than use only their
ratios.
The most significant constraints that we have derived are
those on $\lambda'_{222}$, $\lambda'_{223}$ and $\lambda'_{31k}$.
All of these bounds are {\em new} from the phenomenological
standpoint. With this analysis, only one set of
lepton--number violating \rp\ couplings, namely
$\lambda'_{32k}$ remains relatively unconstrained. This
gap would, most likely, be difficult to fill from low--energy
experiments, with rare
$D$-decays perhaps being the best hope!

\noindent
We thank S.~Banerjee for useful discussions.

\end{document}